# Rediscovering Ranganathan: A Prismatic View of His Life through the Knowledge Graph Spectrum

**Biswanath Dutta[1] and Shakeeb Arzoo[2]**

**[1,2] Documentation Research and Training Centre, Indian Statistical Institute, 8th Mile Mysore Road, RVCE post, Bangalore-560059.**

**[1]bisu@drtc.isibang.ac.in; [2]arzoo@drtc.isibang.ac.in**

The present study puts forward a novel biographical knowledge graph (KG) on S. R. Ranganathan, one of the pioneering figures in the LIS domain. It has been found that most of the relevant facts about Ranganathan exist in a variety of resources (e.g., books, essays, journal articles, websites, blogs, etc.), offering information in a fragmented and piecemeal way. With this dedicated KG (henceforth known as RKG), we hope to furnish a 360° view of his life and achievements. To the best of our knowledge, such a dedicated representation is unparalleled in its scope and coverage: using state-of-the-art technology for anyone to openly access, use/re-use, and contribute. Inspired by Ranganathan's theories and ideas, the RKG was developed using a "facet-based methodology" at two levels: in the identification of the vital biographical aspects and the development of the ontological model. Finally, with this study, we call for a community-driven effort to enhance the RKG and pay homage to the Father of Library Science on the hundredth anniversary of his entry into the field of librarianship, which he revolutionized and revitalized as it developed into Library and Information Science.

**Keywords:** *Knowledge Graphs, Linked Data, Ontologies, Biographies, Ranganathan S. R., Artificial Intelligence, ChatGPT, Applications, Methodology*

## Introduction

The art of biography is an ancient one, going back to Plutarch's *Lives*. Biographies have always been a highly sought-after information resource— as we are innately drawn to the lives of other people. Whether it be an article entry in a biographical dictionary, a chronological timeline breaking down the major milestones of an eminent personality, a book-length treatment of an individual, a life spanning across several volumes (like Leon Edel's multi-volume work on Henry James[1]), structured (audio/video or textual) interviews that capture a person's vitality, or a website compiling all of the above, the lives and experiences of other people never cease to fascinate us.

The *Oxford English Dictionary*[2] puts forward the definition of biography as "somebody's life story written by somebody else." A more comprehensive definition is given by *Encyclopaedia Britannica*[3] as "a form of literature, commonly considered non-fictional, the subject of which is the life of an individual. [It is] one of the oldest forms of literary expression [and] it seeks to re-create in words the life of a human being—as understood from the historical or personal perspective of the author—by drawing upon all available evidence, including that retained in memory as well as written, oral, and pictorial material." Going by the descriptions quoted above, we can say that biographies attempt to reconstruct an individual's life in the context of their time. They attempt to bring out the biographee's personality, social networks, major milestones, influences and impact, and attempt to explain the entire life-story from varying perspectives. Some of these "aspects" is discussed later on in this paper.

The primary problem in these documents stems from their "free-text" nature. Apart from being "machine-unfriendly", they put a certain cognitive toll on humans using them. Although biographies often annotate their text extensively by means of footnotes, back-of-the-book notes, chronological tables, indexes, and references, they are still by and large unstructured. The very textual and bound nature of the documents does not allow for free exploration of the many entities (related individuals, organizations, ideas, terminology, etc.) that are present in biographies. One of the advances made possible by the Web was the idea of hyperlinks, through which people could jump back and forth between the relevant resources. Most biographical databases (e.g., Biography Online, Standford Online) utilize such techniques unsparingly (through threading, infoboxes, images, references, etc.), presenting the biographical information in a multifaceted manner. However, the data present is not particularly queryable amenable to analytics, or "semantically enriched" for machine processing.

Knowledge Graph (KG) offers a way to represent and store data within a graph database. It models the underlying data as a graph ("nodes-and-edges"), transforming the unstructured data into a structured set of data embedded within the framework of a rich scheme that makes implicit assumptions explicit and provides a 360º view of a phenomenon of interest. It thus becomes a powerful tool for knowledge discovery and a platform to explore further connections. In the parlance of Linked Data principles, it lets us look up more things with the help of already existing, known things[4]. Apart from these, KG makes the data amenable to intelligent analytics, enables semantic search, allows for the inferencing of new knowledge, serves as a foundation for AI applications, and so on. Modelling data as a graph is a step ahead of relational databases[5] as KGs are better equipped to handle data that are incomplete or missing[5]. Because KGs are painstakingly modelled with entities clearly disambiguated and contextually enriched, they form a part of explainable AI that is not plagued by the issues running in Large Language Models (LLMs) like ChatGPT and PaLM[6]. For example, if we run the query: *Provide the details of S. R. Ranganathan's conjugal life*; the model either provides incorrect responses or is silent (Figure 1).

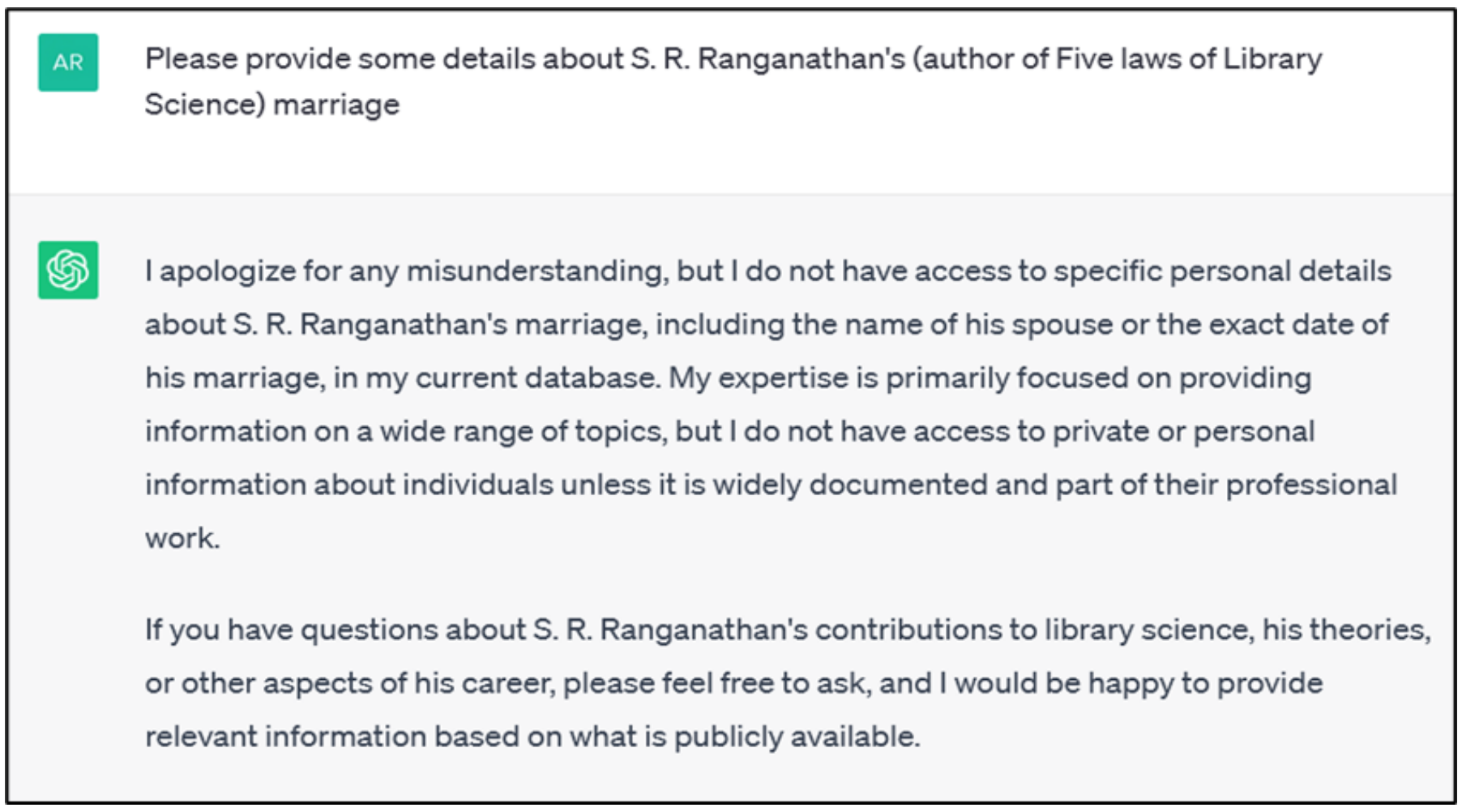

Fig. 1 - ChatGPT's response to a query about Ranganathan's marriage

If we follow the prompts and ask for his employment details, the model answers it as shown in Figure 2:

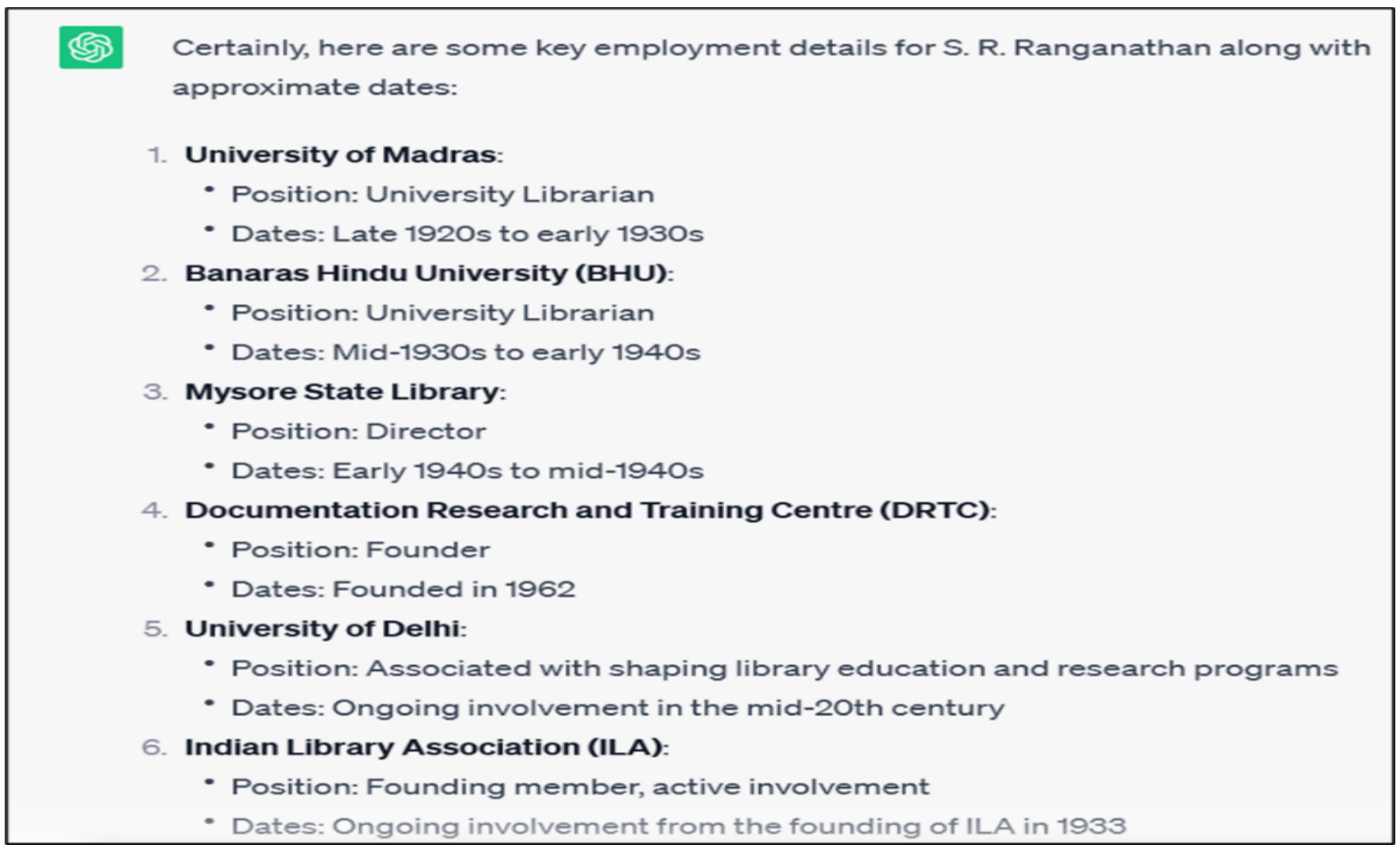

Fig. 2 - ChatGPT attempting to answer a query about Ranganathan's employment

The details are only partially correct. The correct information can, of course, be gleaned from the various biographical sources listed in this paper, or by querying our data model. This is because we have carefully identified, curated, and populated our KG with the biographical data (see section on "Design of RKG: Methodology and Results" for a detailed breakdown). To be fair, we cannot expect LLMs to respond with such exactitude as they have a different cross-domain focus and a voluminous data bank. But carefully engineered KGs can rise to such challenges and offer a plethora of benefits. KGs thus offer a semantic understanding of a domain because they contextualize the knowledge in a structured manner, allowing us to define categories, their relationships, and the individuals involved. The carefully curated data then allows for precise querying and the ability to infer more knowledge from existing data[5,7]. They also facilitate data integration from diverse sources and scale accordingly.

S. R. Ranganathan is a renowned figure in the LIS domain. He is popularly known as the "Father of Library Science" and his birthday is celebrated as National Librarians' Day in India. His achievements in this field are unparalleled— from the Five Laws of Library Science to a new system of classification, various theories and terminologies, etc. Despite this, there is no common platform to find the necessary information on Ranganathan. They are scattered across various articles, books, websites, and encyclopaedias. One of the main contributions of this work is to assist researchers, librarians, students, and other curious minds in their search for authentic sources and information. We attempt to systematically unpack his life, covering all major or minor biographical details, provide a consolidated bibliographic list, and make the data amenable to analytics. To the best of our knowledge, such a dedicated representation is unique in its scope and coverage: using state-of-the-art technology with open standards for anyone to access, reuse, contribute, and further enhance.

The major objectives of this undertaking are:

1. A dedicated KG that presents carefully curated biographical and bibliographical data related to Ranganathan.
2. A common platform to search for information about S. R. Ranganathan that anybody can contribute to and further develop.
3. To pay homage to the Father of Library Science in India in our capacity as LIS professionals and to issue a call for a community-driven effort to promote further research into the life and works of S. R. Ranganathan.

The contributions discussed in this study are:

1. A comparative review of related efforts.
2. A discussion of the Ranganathan Knowledge Graph (RKG) along with an overview of the biographical ontology "Ontobio" that was specifically designed for this work.
3. A review of the major critical resources related to Ranganathan utilized in the construction of the knowledge graph.

The rest of the paper is organized as follows: the *Related Work* section conducts a literature review of similar efforts; a section on *Design of the Ranganathan Knowledge Graph: Methodology and Results* discusses the various phases in the development of the KG, from the identification of the relevant resources, the description of the biographical ontology, the methods used in populating the ontology, thereby transforming it into a KG, and a semantic evaluation of the model. Finally, the last section discusses the results, limitations, and the scope for future work.

## Related Work

Information resources about a person exist either in the form of textual or web documents. As mentioned in the *Introduction,* the idea of utilizing KG technology to describe people lies in drawing rich relationships among the entities implicated in their lives, such as people, organizations, places, the various roles and activities undertaken by them, their particular cultural circumstances, their achievements, etc., in such a way that machines can meaningfully manipulate the knowledge for inferencing and analytics. In a sense, resume- or CV-type ontologies, as well as the highly popular FOAF vocabulary, have been proposed to represent some biographical information about people (further detailed in the succeeding paragraphs). Besides that, there are some dedicated biographical ontologies (formal and informal) that attempt to capture biographical information in a machine-processable manner.

Web resources that are more traditional in their content and presentation (simple HTML pages) convey ample biographical information about eminent figures. They usually depend on a community of contributors to develop their resources (like Wikipedia). They are then peer-reviewed before being made public. For example, The Nabokovian[8] hosts a variety of resources on the preeminent Russo-American novelist of the last century Vladimir Nabokov, that range from a biographical timeline, impact and influences, to annotations to major works, bibliographical resources, etc. Similarly, the Dutch Kafka Circle[9] aims to promote the study of the life, times, and works of Franz Kafka in the Netherlands. It publishes articles, news, and reviews; organizes lectures and study circles; and sponsors other activities. Apart from these dedicated websites, encyclopaedic resources contain curated information on famous personalities throughout history. For example, the well-respected *Encyclopaedia Britannica*[3] commissions articles from eminent scholars to provide versatile information (including references to other works) on varied personalities. In the same vein, the Stanford Encyclopaedia of Philosophy[10] endeavours to present the best bio-critical articles from specialist scholars on various Western philosophers. Biography Online[11] and Wikipedia are a few of the other free online resources that maintain similar information. Several publishing giants, like Penguin, Oxford Books, and Harper Collins, also maintain biographical and bibliographical information on most of their top-selling authors.

KGs representing biographical information about eminent personalities start with the Google Knowledge Graph. As we know, Google search is based on "things, not strings[12]", a system made possible due to Google acquiring and maintaining Freebase and Wikidata[13]. Both of these knowledge bases contain highly structured data on people, world events, and scientific information, and are encyclopaedic in scope. Ranganathan, for example, is given the identifier Q457933[14] in Wikidata and is described with a host of assertions and properties.

Hyvönen et al. (2019)[15] have developed the Semantic National Biography of Finland (BiographySampo), which would augment the reading experience of biographical texts such as those found in dictionaries and encyclopaedias with the help of Semantic Web Technologies. They build on their previous work, BIO-CRM[16], which is a domain specific extension of the ISO standard CIDOC-CRM, furthering the cause of prosopographical research. This KG automatically creates life-stories for people through the use of NLP tools from heterogeneous sources which, on search and retrieval, bring back Persons, Places, Social Networks, Life Maps, and Relations in a database comprising some 13,000 biographical entities. They are further linked to external datasets to enrich the portal for public use.

The EventKG+BT was proposed as "a system to explore personal biographies through various timelines[17]". It builds upon the authors' earlier work[18] where they formalized "a temporal knowledge graph" (Event KG) that integrated data existing in heterogeneous schemas in an RDF form. Using such a KG, it would be possible to generate biographical timelines for all significant events related to famous individuals. The authors maintain a live portal demonstrating their results for famous American presidents.

As mentioned before, there have been dedicated ontologies that take a graph-based approach to representing biographical knowledge. The first such vocabulary to encode the changing states in a person's life was BIO[19]. Following this up, Ramos (2009)[20] took an event-centric approach to the collective biography of people in the Biography Light Ontology (BLO). This model mainly seeks to answer the basic questions of "what, who, where, and when" of people's lives. Ontolife as proposed by Kargioti et al. (2009)[21], is an agent-centric biographical model that touts itself as "semantically managing the personal information of people." It bears the most resemblance to our undertaking, as it is built upon the Person entity with the rest of the relationships and properties asserted around it. Its scope describes a person's relationships, biological and demographical descriptors, education, qualifications, periods, events, and so on.

Apart from these dedicated biographical ontologies, there are certain related ontologies that represent biographical information, such as the FOAF vocabulary[22] (describing people along with their networks and interests), the VIVO ontology[23] (describing scholarly activities for individuals and organizations), and CV-type ontologies (describing a person's qualifications, skills, basic biographical data, etc.) such as ResumeRDF[24], etc., that can be used to create a "patch ontology" engendering a description of the Person class. For the development of RKG, we developed the Ontobio ontology, which is described in the "Biographical Ontology Ontobio" section below. This biographical ontology overcomes the limitations of previous ontologies and comprehensively covers all the facets of an individual's life. Furthermore, it is an integrative framework that supports the re-use of existing vocabularies. We will address some of the features later on in this paper.

## Design of RKG: Methodology and Results

This section discusses the phase-wise development of RKG. The first phase, "Resources and Associated Challenges", discusses the identification and collection of biographical and bibliographical data related to Ranganathan. The next phase, "Biographical Ontology Ontobio", where we describe the biographical ontology, provides the structural framework for the KG. The succeeding section, "Knowledge Graph Realization", describes the development of RKG starting from data selection and analysis to their eventual integration into the ontology forming the KG. The developed RKG is further evaluated through two distinct approaches provided in the "Evaluation of the Model" section.

### *Resources and Associated Challenges*

A major aspect of biographical research is the identification of authentic sources that shed light on the biographies. For a biographer, it involves the unravelling of the "paper trail" left by the biographical subject. In our research to retrieve the biographical facts of Ranganathan, we mostly consulted sources that are purely secondary sources of information. We have not examined any first-hand biographical sources. However, our documentary sources for biographical information are from people directly connected to Ranganathan, e.g., his son: Yogeshwar Ranganathan; his students and friends: M. A. Gopinath, Bernard Palmer, etc.; his contemporaries: Eugene Garfield, R. N. Sharma, etc.

We started our information search with Wikipedia, encyclopedia entries, and various databases containing authoritative information. The article on Ranganathan in the Encyclopedia of Library and Information science by M. A. Gopinath[25], a close associate of Ranganathan, was of much use. The next major resource was the biographical study by Yogeshwar Ranganathan[26], which provided a rich context for his father's life, environment, and achievements. M.P. Satija's review article[27] in International Library Review is a valuable document in that it singles out all research related to Ranganathan. Since most of the information was in a physical format, we had to prepare multiple documents covering the various aspects of his life and work. A list of all resources relevant to this study is provided in Table 1. The bibliographical details related to Ranganathan were available from ISKO (edited previously by the scholars Raghavan[28] and Bianchini[29]) and had to be web-scraped using Python's Beautiful Soup module. The resulting Excel file was later cleaned using various data manipulation functions and cross-verified using external databases (for details see the section "Knowledge Graph Realization" below). Throughout this undertaking, we stored and maintained the data in multiple Word documents and spreadsheets, using Google Cloud Storage for backup. Maintaining multiple documents was necessary for revising, editing, and updating the factual data belonging to different categories (e.g., dressing style, food habits, diseases) distinctly. We provide an example of the data collection related to Ranganathan's food habits in Figure 10.

One of the major obstacles to the collection and curation of biographical data is the unavailability of important publications. Some of them are out-of-print[30], while others are not yet available in a digitized form[26] (and are very hard to locate) that would make the progress easier. We hope that the research conducted during this work will serve as a starting point for scholars, students, and other professionals to collaboratively enhance the KG presented herein (similar to the Wikidata knowledge base). As mentioned previously, a paucity of information on Ranganathan in Wikidata[14] and DBPedia[31] was observed. We hope that our linked, structured data model will further enrich those knowledge bases. Furthermore, we also envision that the present work will also support AI applications, like ChatGPT for rich and accurate information retrieval about Ranganathan.

The second challenge revolved around the data model. As we hinted in the *Related Work* section, we didn't come across an ontological model that concentrated exclusively on an individual, bringing forth personal experiences, cultural background, personality, and achievements. Some of the biographical concepts (such as daily routine, beliefs, and personality aspects) that are easily understood by readers can be surprisingly tricky to model. This led to the creation of a biographical ontology “Ontobio” which would not only serve as an integrative model for previous efforts but would also be flexible and robust enough to cover all the biographical aspects of a person comprehensively. Finally, we know that writing SPARQL queries to retrieve the necessary data from the KG can be quite challenging for laypeople. In future work, we aim to create a live SPARQL endpoint with clear documentation that would assist users in retrieving the necessary information (see Conclusion for such proposals).

We conclude our discussion of this phase of development of the RKG by providing a complete list of the biographical, bibliographical, and critical resources relating to Ranganathan (Table 1). In the next section, we discuss the ontology that provided the scaffolding for the KG.

Table 1. A List of Important Resources necessary for this Knowledge Graph

| Sl. No. | Bibliographic Details | Comment |
|---|---|---|
| **1.** | Gopinath, M. A. (1978). Ranganathan, Shiyali Ramamrita. Kent, A., ed. *Encyclopedia of Library and Information Science*. New York: Marcel Dekker. | A bio-critical essay listing all major life and professional events of Ranganathan. |
| **2.** | Gopinath, M. A. (1994). Memorabilia Ranganathan. *Ranganathan Centenary Series*, 5. Bangalore: Sarada Ranganathan Endowment for Library Science. | Same as resource no. 1, with some minor modifications. |
| **3.** | Yogeshwar, R. (2001). *S. R. Ranganathan, Pragmatic Philosopher of Information Science: A Personal Biography*. Mumbai: Bharatiya Vidya Bhavan. | A major resource attempting to capture the life and times of Ranganathan. |
| **4.** | Palmer, B. (1976, 1977). Encounter with Ranganathan (In three parts). *Herald of Library Science*, 15(3-4), 339-41; 16(2-3), 215-19; 16(4), 392-98. | An important essay detailing personal recollections with critical comments. |
| **5.** | Garfield, E. (1984). A Tribute to S. R. Ranganathan, the Father of Indian Library Science. *Essays of an Information Scientist*, 263-79. | An essay paying homage to Ranganathan’s achievements. |
| **6.** | Satija, M. P. (1987). Sources of Research on Ranganathan. *International Library Review*, 19(3), 311-20 | A starting point for scholars embarking research on Ranganathan. |
| **7.** | Satija, M. P. (2022). From stammering to eloquence: Ranganathan as a speaker and orator. *Annals of Library and Information Science*, 69(2), 169-172. | An evaluative account of Ranganathan’s personal struggles. |
| **8.** | Sharma, R. N. (1979). S. R. Ranganathan: A Personal Tribute. *The Journal of Library History* (1974-1987), 14(1), 58-72. | A tribute to Ranganathan’s professional achievements. |
| **9.** | Raghavan, K. S. (2019). Shiyali Ramamrita Ranganathan. *International Society for Knowledge Organization (ISKO)*. (Retrieved March 10, 2023 from https://www.isko.org/cyclo/ranganathan) | A bio-critical account of Ranganathan’s life and works. |
| **10.** | Bianchini, C. (n. d.) Bibliography of Ranganathan's works. *International Society for Knowledge Organization (ISKO)*. (Retrieved March 10, 2023 from https://www.isko.org/cyclo/ranganathan#bib) | A web resource based on a more comprehensive bibliographical resource; very useful for quickly parsing through the data. |
| **11.** | Dasgupta, A. K. (1965). *Ranganathan: An Essay in Personal Bibliography: A Bibliography on and by Ranganathan*. Bombay: Asia Publication House. | A unique resource capturing all of Ranganathan's publications till 1961. |
| **12.** | Weinberger, D. (2012). *Rediscovering Ranganathan*. (Retrieved May 25, 2023 from https://semanticstudios.com/pdfs/forrester.pdf) | A tribute to the long-lasting ideas of Ranganathan that are still relevant to the workings of Internet. |

***Biographical Ontology Ontobio***

Ontologies are known as "foundation layers for knowledge graphs[32]". In other words, they form the skeleton structure of a KG, encapsulating the data within a tailored schema. As discussed previously (in section on "*Related Work*"), existing ontologies were found lacking when it came to comprehensively modelling the varied biographical aspects of a person. Their vocabularies were either too restrictive or the links (URIs) were not dereferenceable. Furthermore, no attempt was made to model a person's personality traits, social relationships, hobbies, travel details, associated places and organizations, awards, professional titles, etc., within a single framework. Their purpose was limited to answering the basic "what, why, when, and where" of people's lives. In contrast, we required a model that could cover all the major aspects of the life of an eminent individual (a faceted deconstruction of biographical knowledge is presented below). In a manner inspired by Ranganathan's theory of faceted classification, we identified the major facets of an individual's life that are usually presented as biographical information and broke them down to their core constituents. This resulted in a person-centric ontology through which the social networks, places frequented, temporal instances, achievements, etc., could be clearly represented. It should be noted that in conformance with best practices, we reused existing vocabularies. Some of the common concepts, such as Places and Gender were used from Schema.org[33], while Agents and Organizations were reused from the FOAF vocabulary[22], and the BIBO ontology[34] and the AMV vocabulary[35] were put to use for the publication details. The BIO vocabulary[19] was used for concepts such as Marriage and Employment as was ResumeRDF[24] for other professional aspects.

The development methodology for Ontobio followed the YAMO methodology[36], which in itself was inspired by Ranganathan's theory of analytico-synthetic classification. At the core of this methodology lies certain of Ranganathan's guiding principles (as postulated in his *Prolegomena to Library Classification*), namely the canons of relevance, ascertainability, context, currency, exhaustiveness and exclusiveness, permanence, reticence, etc. The reader is referred to a previous work[36,37] for a detailed exposition of the ontology development methodology and the principles behind them.

The Ontobio ontology is based on three fundamental aspects. As discussed in the preceding sections, we consulted multiple biographical resources in the form of biographical indexes, vocabularies, books, and encyclopaedic articles, to understand the domain. It was found that in the biographical literature relating to distinguished authors and academicians[38, [39], three major aspects are frequently commented upon. They are *Life and Character*, *Work and Thought*, and *Publications*. For the purposes of our present study, we generalize them as *Personality*, *Environment and Milieu*, and *Achievements/Milestones*. We further break down the three aspects and itemize the concepts within them that are usually found as biographical data (see Figures 3, 4 & 5). These aspects form the inner casing for the RKG, which can be further extended to accommodate the biographical data of other individuals.

*Personality of the Subject*: This refers to the individual characteristics, including psychological and social traits, that are partly a result of nature (idiosyncrasies that are a result of genetic inheritance and a reaction to one's upbringing) and nurture (the particular circumstances in which an individual has been brought up). We have attempted to identify some of the facets that bring out the "personality" of an individual, as shown in Figure 3. Some of the elements (like *Gender* and *Physical Characteristics* like height, and body size) are biologically determined, but they are too individualistic to be described anywhere else.

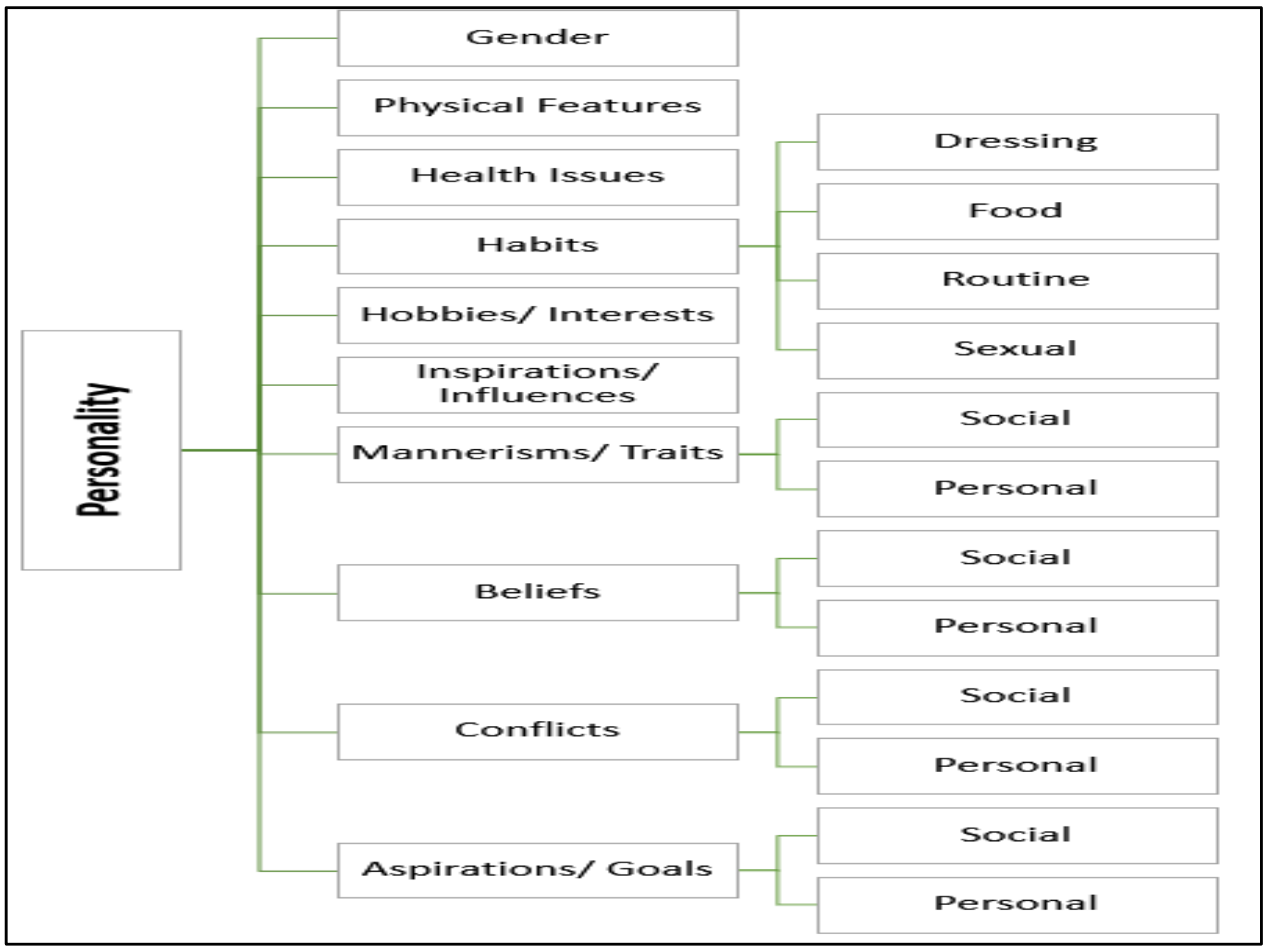

Fig. 3 - Personality Aspects

Personality aspects such as *Aspirations* are those goals that a person sets for him- or herself (they may be Family or Professional) while *Inspirations and Influences* (they may be in the form of books or other individuals) are those aspects that leave an indelible mark on an individual's psychological make-up. His or her *clothing* and *attire*, his or her *food habits* (whether vegetarian or non-vegetarian), and his or her *regular routine* (such as his or her *morning walks, dining time*, and *working hours*), and his or her beliefs whether privately (i.e., personal) held (e.g., *Communism is good*) or those publicly (i.e., social) expressed (e.g., *Religion fortifies one's character*), all form an essential part of their *Personality.*

*Environment and Milieu*: The social environment plays a great role in moulding and shaping a person's life. As mentioned above, many of the personality facets are a result of the particular circumstances and the values of the social groups[40] that the individual absorbs and integrates within his or her personality. Also of special mention is the Milieu in which the individual lives which influences his or her *dressing*, *food*, and *social habits*. The facets that make for the "surroundings" for the individual are shown in Figure 4.

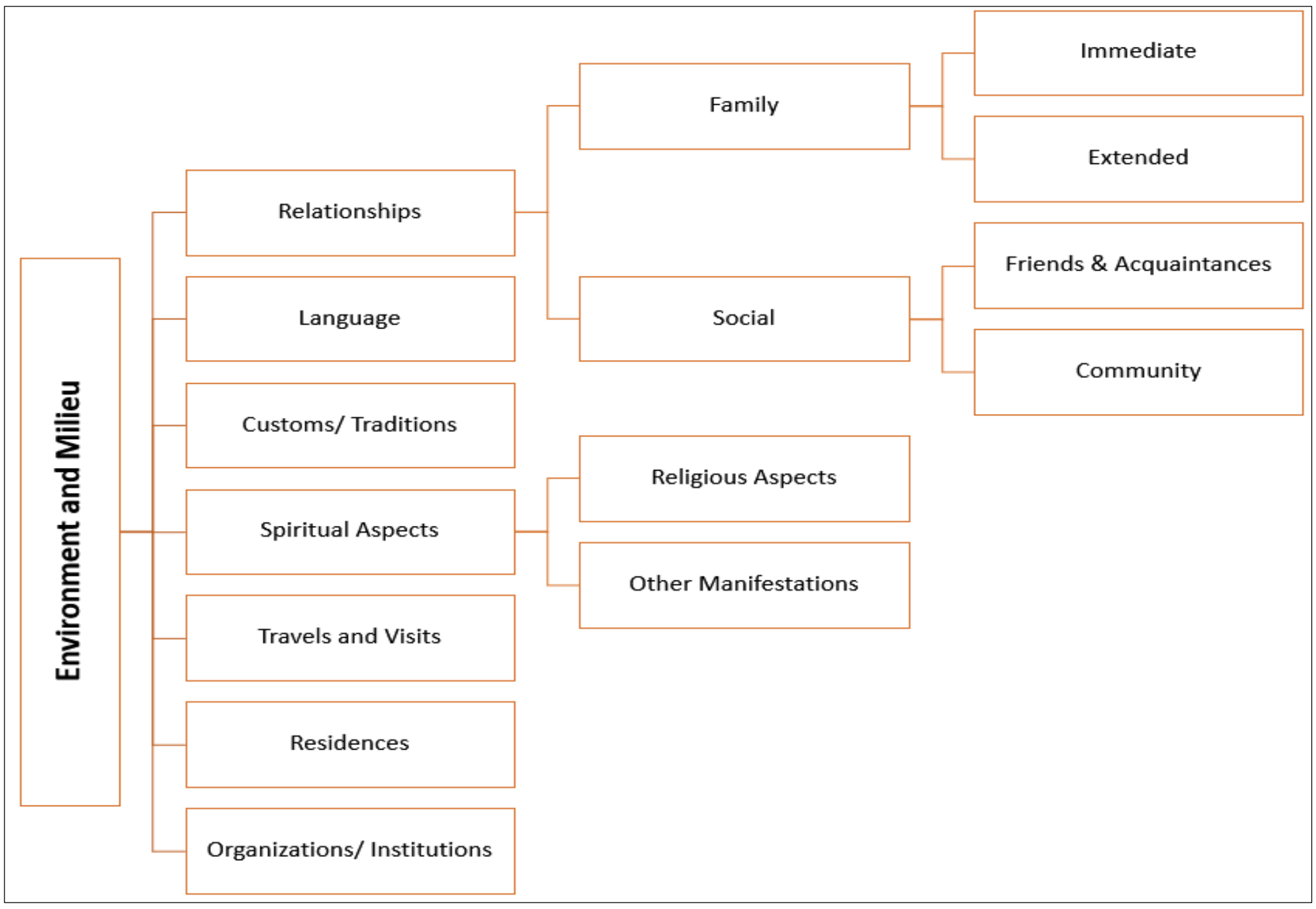

Fig. 4 - Aspects from the Environment and Milieu

As Figure 4 shows, the aspect *Environment and Milieu* is composed of several components and sub-components. *Family Relationships* are divided into two: immediate, which comprises parents, siblings, uncles, and aunts; and extended, which refers to cousins and in-laws. Furthermore, social relationships consist of friends, acquaintances, students, teachers, and the community to which an individual belongs (e.g., South Indian Ayyars, Gujaratis, Bengalis, etc.). A person usually participates in various *religious practices* (puja, namaaz), *language* (Tamil, English, Hindi, etc.), and *customs*, as a result of his or her particular locale, while the various organizational affiliations and their changing residences form a backdrop for the person's individual undertakings.

*Achievements/ Milestones*: The subject of a biography is usually a distinguished individual. Besides noting down his or her struggles and experiences, how the particular achievements were brought about is of great interest to the biographer as well as the readers. From the examination of biographical literature, we can conclusively say that the major milestones in an individual's life are the result of the two aspects mentioned before. The facets comprising the *Achievements* in a person's life are shown in Figure 5.

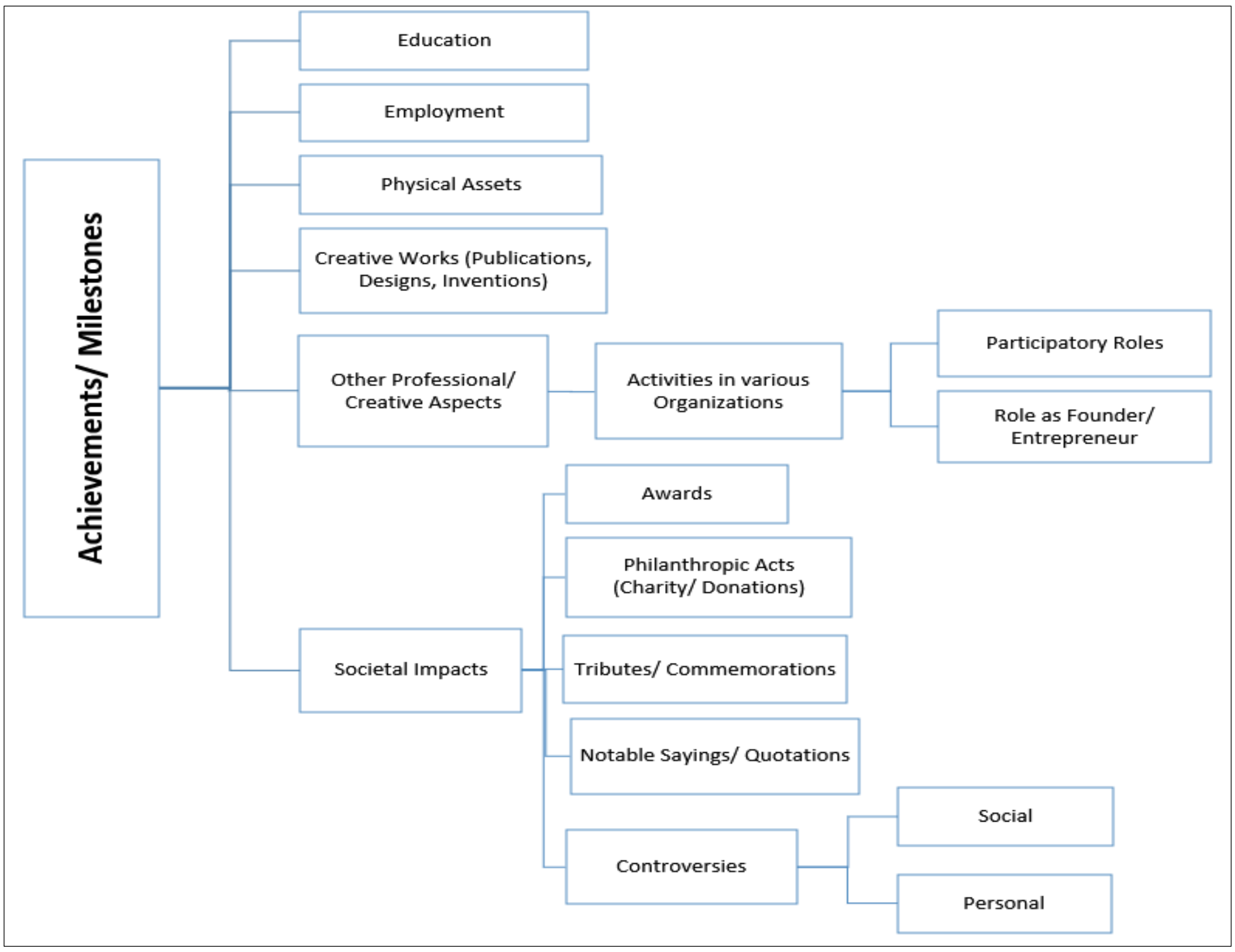

Fig. 5 - Aspects comprising Accomplishments/Milestones

A biography often describes an eminent individual whose societal impacts are visibly manifest. A person's *education* (completed degrees), *employment* and *affiliation* (with various professional bodies: Universities, Government, Private, etc.), and his or her *creative works* (in the form of publications, paintings, or music) all fall under the *Achievements* aspect. Furthermore, the honours bestowed upon an individual by society (such as a Nobel Prize, or a Padmashri award), along with his or her famous (popular or erudite) sayings, fall under this rubric.

The faceted approach to biographical knowledge described above forms the backbone of the Ontobio ontology. After the basic design of the knowledge model has been realized, ontology development necessitates the use of a formal language[36, 37] such as the W3C recommended OWL[41]. The modelling was done on the Protégé ontology editor[42]. The various facets, sub-facets, and the concepts identified above were organized and defined as OWL classes in the ontology. Figure 6 provides a snippet of classes and subclasses from Ontobio. These classes are used to instantiate the entities, such as People, Places, Organizations, Awards, and Communities, etc. in the KG. Besides these classes, Ontobio defined numerous associative properties, both in terms of object properties (a.k.a. relations) and data properties (a.k.a. attributes). The object property (OP) enables linking among entities while the data property enables the description of entities. The properties are defined and associated at the class levels. For example, the OP *hasAward* allows for the linking between the classes *Person* and *Award*. Similarly, data properties allow for the capture and description of an entity, for example, through the DP *awardYear* we can state the year (say 1936, which would represent the value) in which a particular award (say Award01, which is an individual belonging to the Award class) was conferred. Figure 7 provides a lateral snippet of OPs and DPs. Figure 8 provides a snippet of the Ontobio ontology obtained through GraphDB[43] that showcases the modelling example that we discussed here. It is worth noting that in Figure 8, SRRanganathan is an instance of a class Person and eventually it is also an instance of the Agent class which in turn is the superclass to the Person class. Figure 9 provides a graph view of Ontobio classes produced using OntoGraf[44]. The full ontology is available from https://w3id.org/ontobio in different serializations such as RDF/XML, Turtle, JSON LD, and N-Triples. The current version of Ontobio consists of 198 classes, 220 OPs, and 69 DPs.

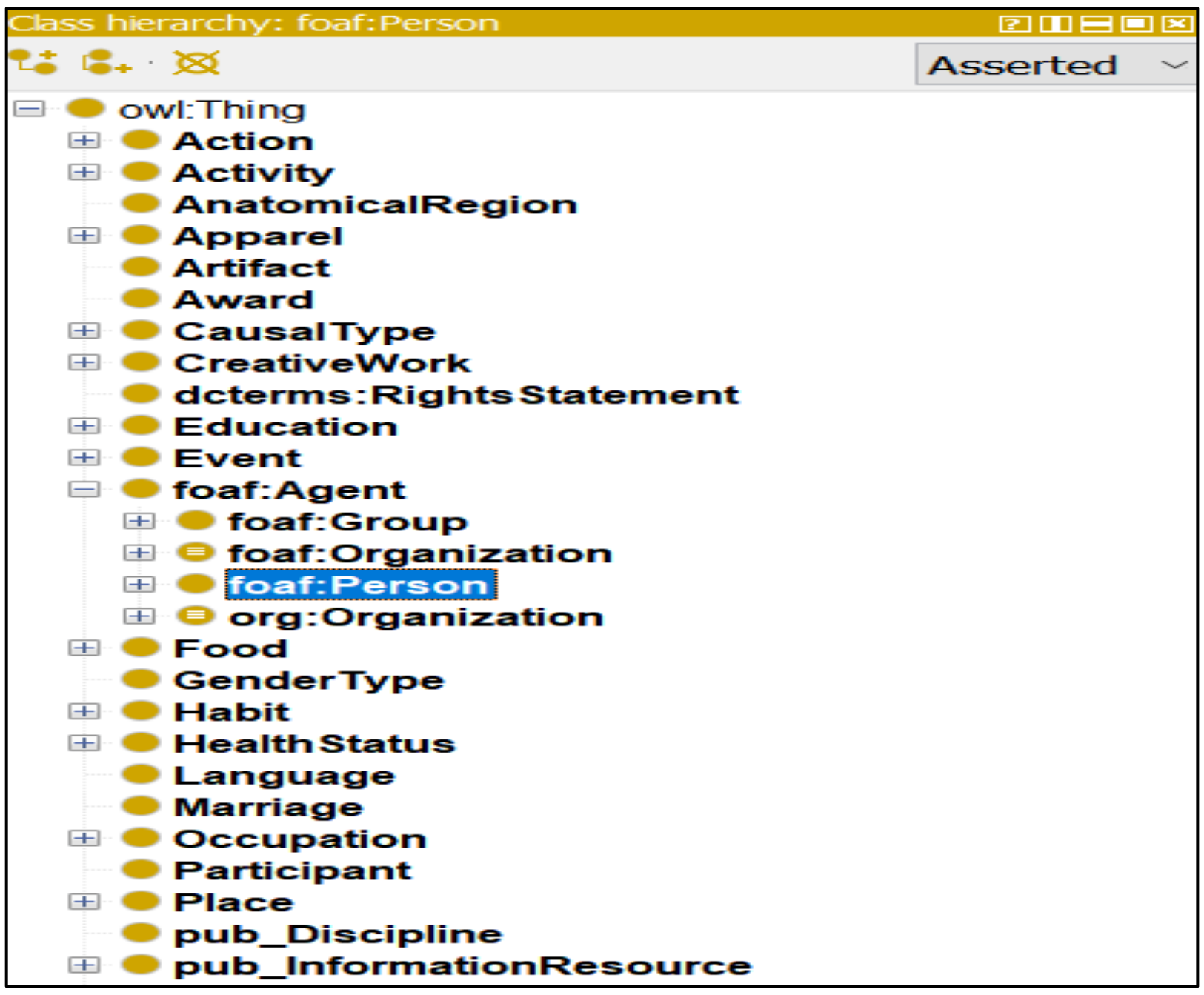

Fig. 6 - A Snippet of the Class Hierarchy from Ontobio

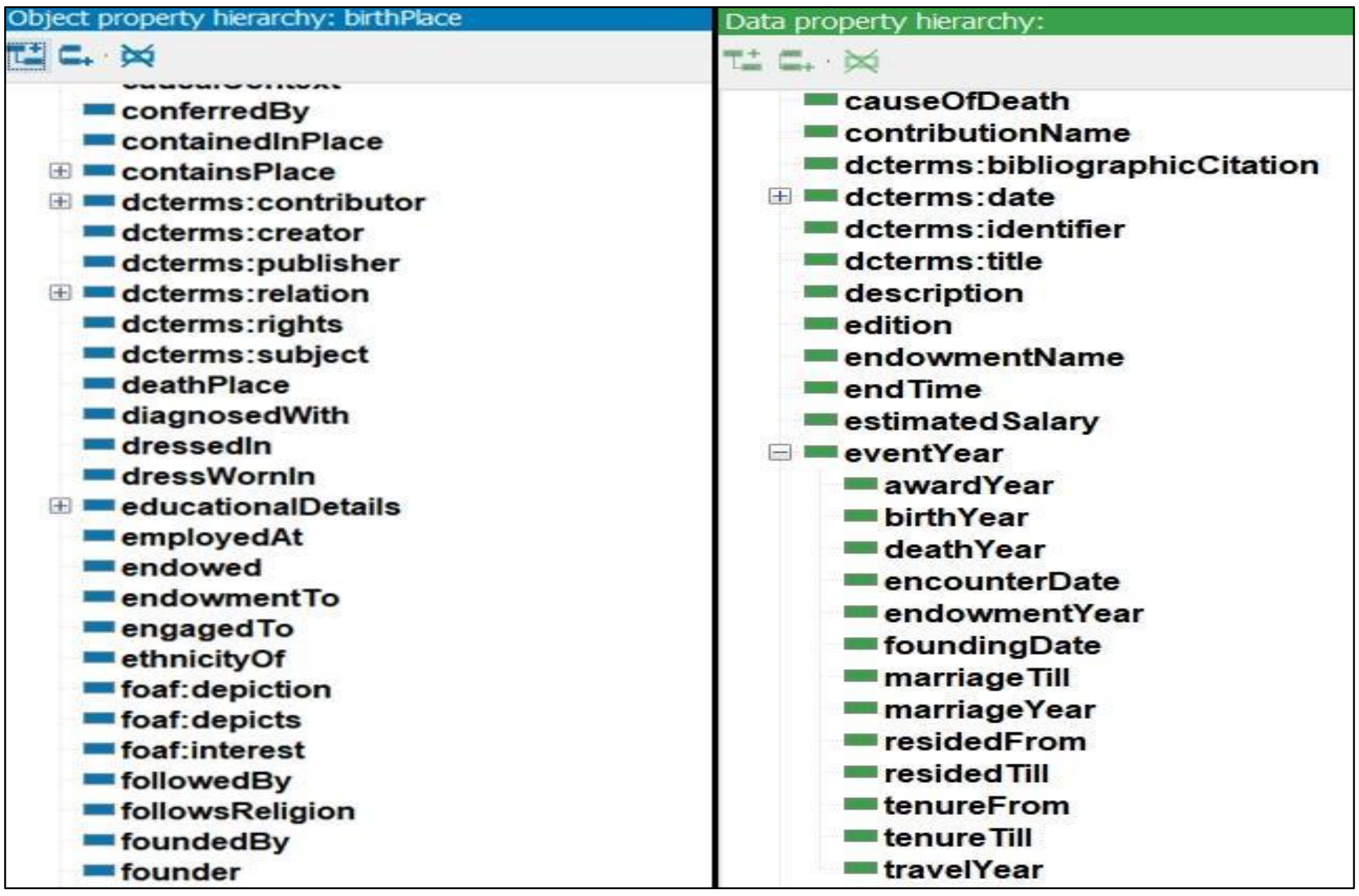

Fig. 7 - A View of the Object and Data Properties from Ontobio

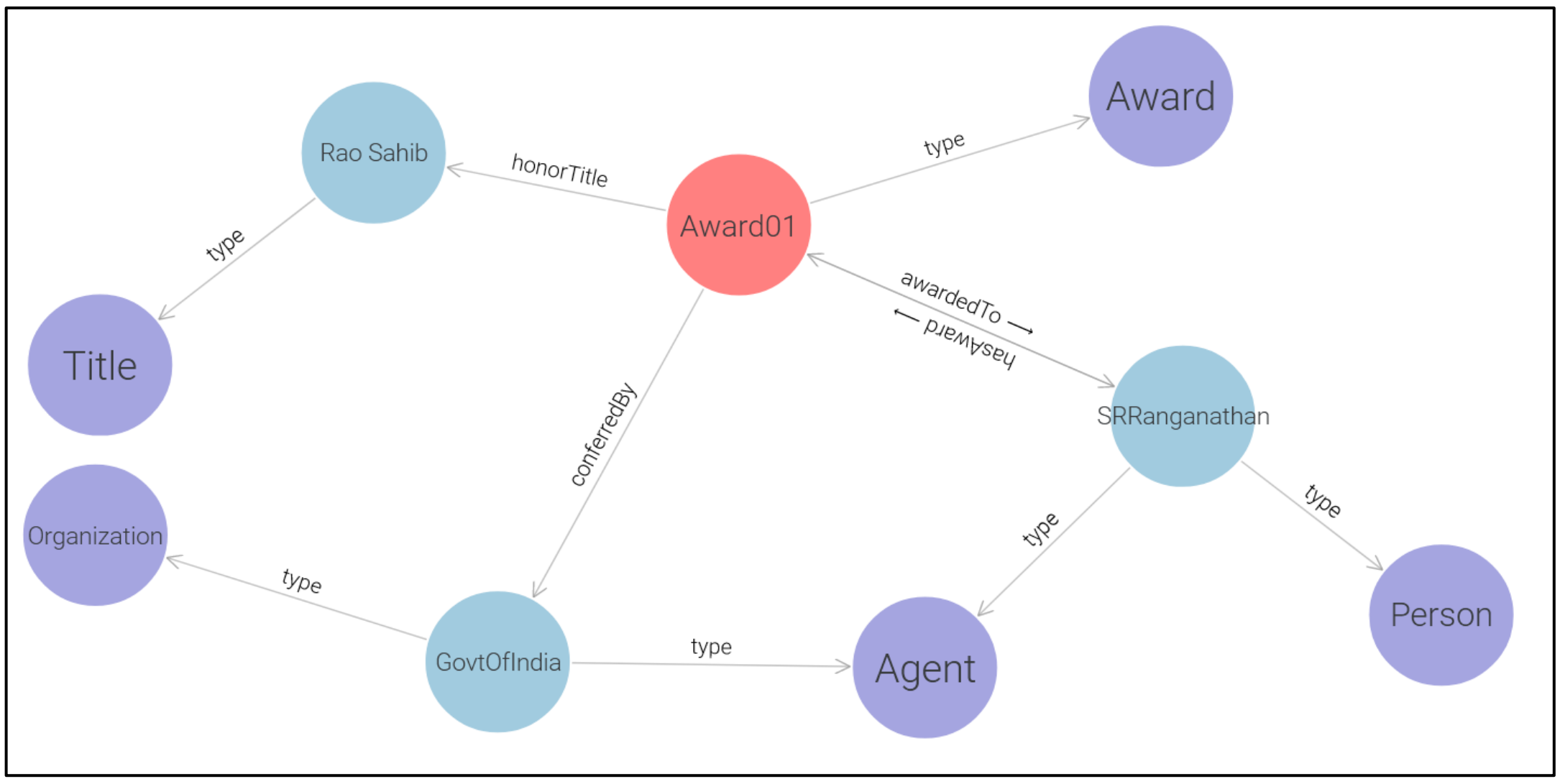

Fig. 8 - A Snippet of Ontobio using GraphDB[43] showcasing a particular modelling aspect. The diagram also visualizes the inferred knowledge, for example, "SRRanganathan" is an Agent. Originally, in the KG, "SRRanganathan" is asserted as an instance of the Person class. Since Person is a subclass of the Agent class (see Figure 6 above), the reasoner inferred the additional information that "SRRanganathan" is an Agent.

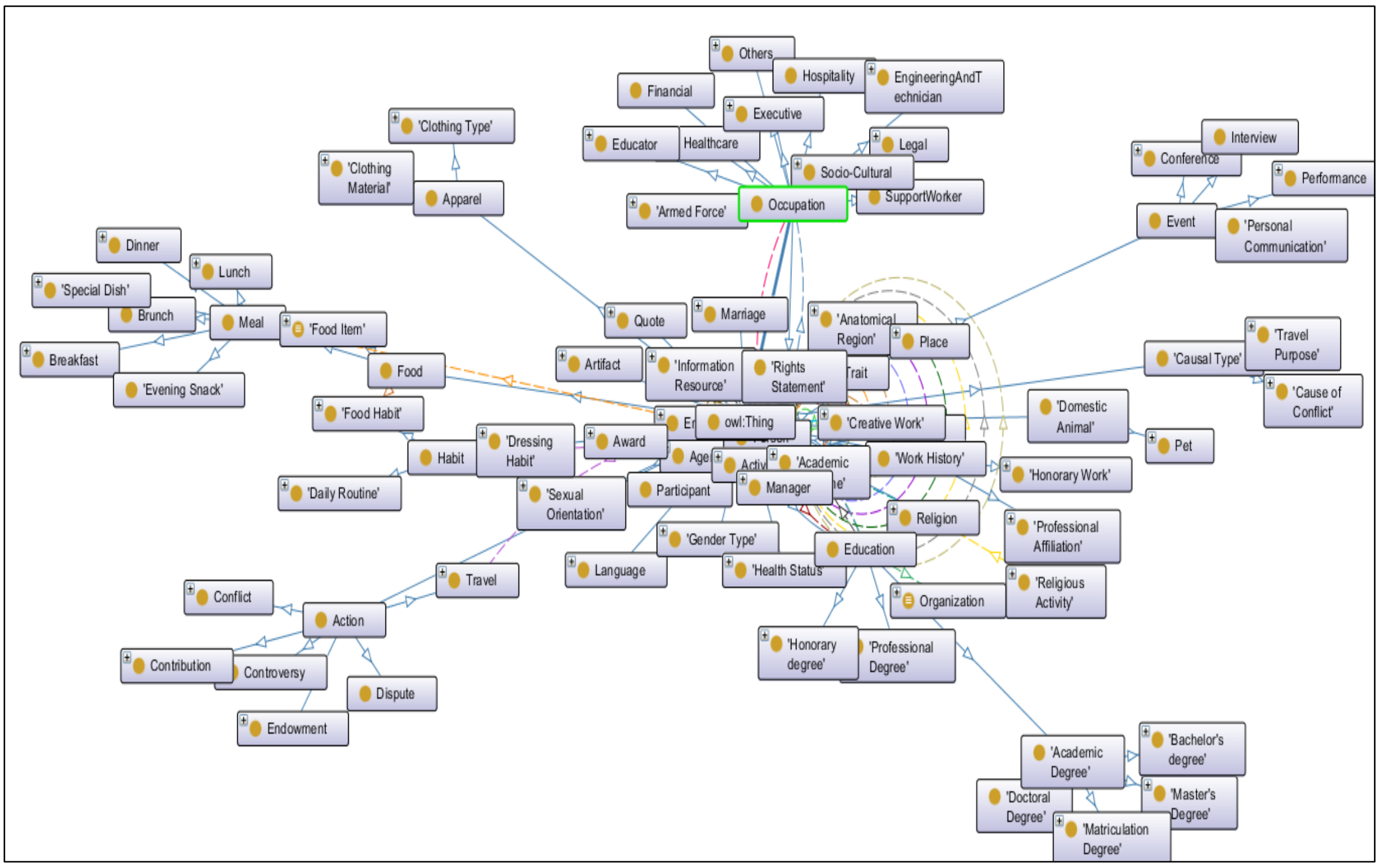

Fig. 9 – Graph view of the Ontobio classes.

The Ontobio ontology can be used to answer questions that people would like to enquire about a given individual, who is, in the context of our project S. R. Ranganathan. In the context of ontologies, such queries are known as competency questions, which help not only with the design and purpose of the ontology but also in its evaluation. Some examples of the questions are provided below. We further discuss some of the queries in a subsequent section on "Evaluation of the Model".

1. What were the nicknames of Ranganathan and his immediate family members?

2. Give us a consolidated list of all the major publications by Ranganathan.
3. Give us the details of the educational degrees achieved by Ranganathan.
4. Name the major awards bestowed on Ranganathan.
5. What were the major health issues befalling Ranganathan?

***Knowledge Graph Realization***

Our research process in the identification and collection of biographical data followed the trajectory discussed in the "Resources and Associated Challenges" section above. Here, we would like to briefly recapitulate the process step-by-step before discussing its effective transformation into graph data.

*Step 1: Identification and Comprehension of the Biographical Data* – The identification and collection of biographical data that was elaborated in the "Resources and Associated Challenges" section formed our first step. Before undertaking any kind of analysis and classification, we had to thoroughly understand the nature of data presented in various sources related to Ranganathan (see also Table 1).

*Step 2: Analysis* – Most of the information found was in free-text form. Our next step was to categorize the heterogeneous data according to its nature. During this process, we had to prepare multiple documents (Food, Travel, Bibliography, etc.) that would cover the various stages of Ranganathan's life, always linking the facts to their original source. Often, we had to verify some facts from the identified sources and resolve the discrepancies (if any). For example, Ranganathan travelled abroad for the first time on a British ship, SS Matiana, on September 1, 1924, in order to get professional education and training in Library Science in London. The voyage began in Washermanpet (Chennai, then Madras in India) and ended in London. This information can be verified through Yogeshwar[26] (pp. 125-6) and Gopinath[25]. A sample screenshot of similar data related to Ranganathan's food habits is shown in Figure 10.

HABITS

**FOOD**

From 1924 *(Yogeshwar p. 138):*

Breakfast: Tapioca Porridge, Bread with milk.
Lunch: Rice, Lentils, Ghee, Vegetable stew, Raita, Paapadam
Dinner: Chapattis, Rice Pudding, Milk and Fruits

From 1925 (*Yogeshwar p. 144*):

Breakfast: Toasted Bread, Wheat flakes, Hot Milk
Lunch: Sambar, Curd, Veg. Curry, Dosa or Upma, Fruits
Evening Snacks: Biscuits and Milk

From 1930s (*Yogeshwar p. 207*):

Morning:

No tea, coffee or any stimulants.
Glass of milk or buttermilk shake with ragi.
Rice and Dal (lentils), followed by Rice and Rasam, Veg. Wafer fries,
then Rice and Curd with pickles.

Afternoon:

Curd-Rice and Idli or Dosa.

Evening:

Rice and Sambar and Rice and Buttermilk.

Special Occasions: (YRB 209)

New rice used to be cooked with a special kind of lentils, seasoned with freshly crushed black

Fig. 10 - Biographical Data of Ranganathan along with the Provenance Information

*Step 3: Data Extraction and Curation* – The bibliography of Ranganathan was available from ISKO in an HTML file. We had to web-scrape it using Python's Beautiful Soup module for its transformation into structured data. The resulting Excel file was later cleaned using various data manipulation functions and cross-verified using the sources identified above. For example, we had to separate the various journal names from their issue and volume numbers, a process carried out through the Text to Columns function in the Data tab in the spreadsheet. Furthermore, misleading or wrong details from the HTML file had to be checked using other resources (see our bibliographical list depicted in Table 1). This took the longest amount of time since the publication output of Ranganathan was huge.

*Step 4: Data Integration* – Finally, the knowledge model, i.e., Ontobio, was populated with data curated in the previous step. This allowed the transformation of our knowledge model into the RKG. The model was populated in two ways: (1) manually entering the data points into the Ontobio ontology using the Protégé[42] interface, and (2) automating it through the Cellfie plugin[45], primarily for the bibliographic data. The latter step involved writing the transformation rules using MappingMaster[46], a domain-specific language for the mapping of spreadsheet content to OWL ontologies. A screenshot of spreadsheet content, here some examples of bibliographic data, is shown in Figure 11; while a screenshot of the transformation rules is shown in Figure 12.

| Title | Year | Journal Name | Issue No. | Notes (Pagination, etc.) | Contributor |
|---|---|---|---|---|---|
| Introduction to the study of character formation | 1916 | Educational Review | v. 22 | 488 | |
| [Mathematical Questions] | 1916 | Journal of the Indian Mathematical Society | v. 9 | Q. 763-64 | |
| [Mathematical Solutions] | 1916 | Journal of the Indian Mathematical Society | v. 8 | Q. 716 | |
| Numbers | 1916 | Educational Review | v. 22 | | |
| Review of Indian mathematics | 1916 | Journal of the Indian Mathematical Society | v. 9 | 6,8 | |
| Homographic sequence | 1917 | Journal of the Indian Mathematical Society | v. 10 | 109-13 | |
| One-one Correspondence | 1918 | Educational Review | v. 24 | 164 | |
| On an Infinite Product | 1919 | Journal of the Indian Mathematical Society | v. 11 | 7 | |
| [Mathematical Solutions] | 1920 | Journal of the Indian Mathematical Society | v. 12 | Q.1090 | |
| Herpholphode Theorem | 1921 | Journal of the Indian Mathematical Society | v. 13 | 87 | |
| Memorandum on Examination | 1922 | Educational Review | v. 28 | 164-78 | Seshu-Aiyar, P. V. |
| Statistical Study of Some Examination Marks | 1923 | Journal of the Indian Mathematical Society | v. 15 | 141-45 | Seshu-Aiyar, P. V. |
| Government contribution in the new provident fund scheme | 1923 | Educational Review | v. 29 | 651-53 | |
| Address at the Pudukottah Library Conference | 1926 | Educational Review | v. 32 | 538-51 | |
| Rural Library Service and National Education | 1928 | South Indian Teacher | v. 1 | 29-35 | |
| A Typical Municipal Library of the West | 1929 | In Diverse Hands: Library Movement | | 143-46 | |
| An Indian library of the eleventh century | 1929 | South Indian Teacher | v. 2 | 261-62 | |

Fig. 11 - Excel Sheet containing Bibliographic Details of Works by Ranganathan

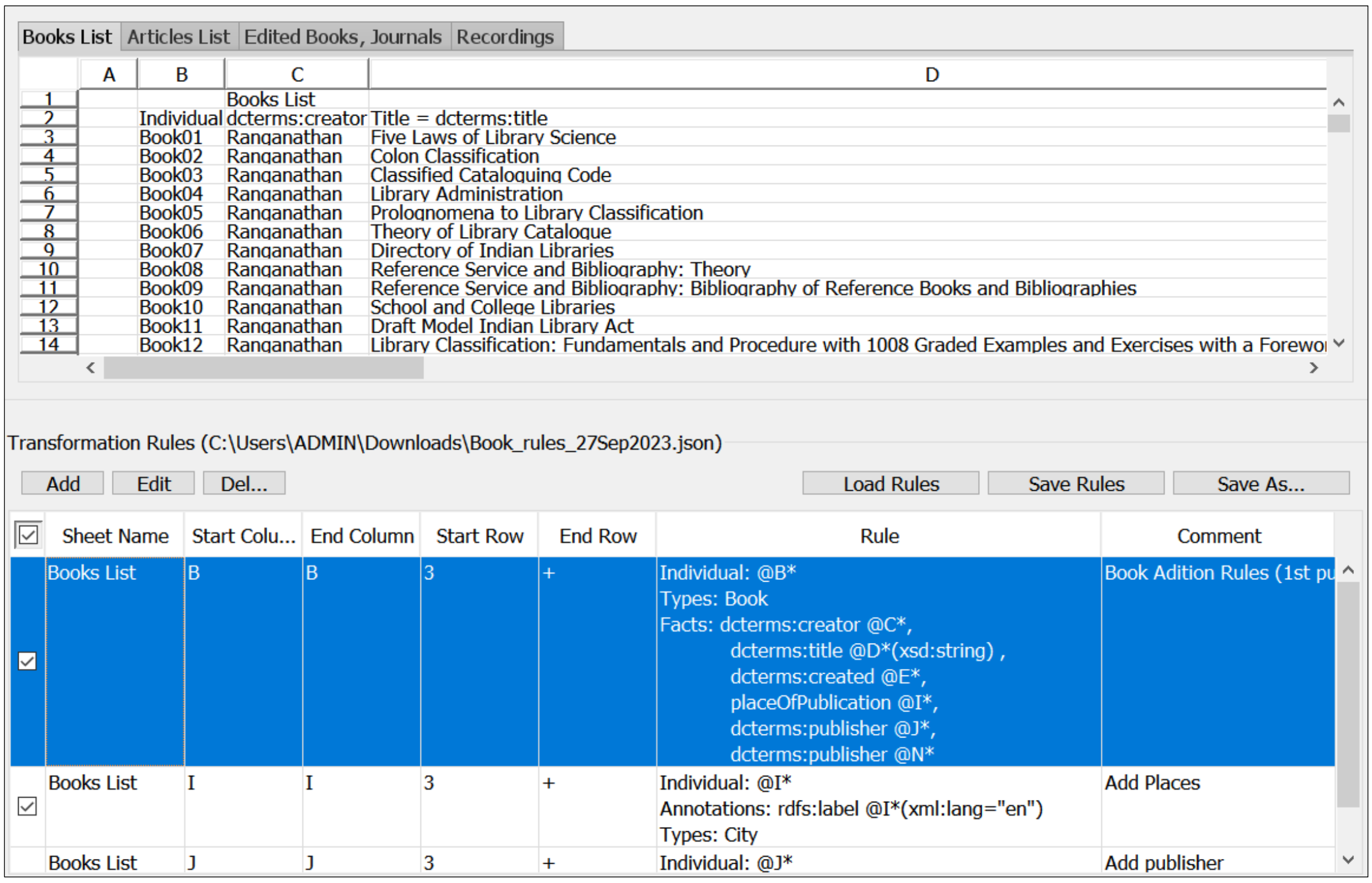

Fig. 12 - Transformation Rules for Ranganathan's Publications using the Cellfie plugin for Protégé

Figure 13 provides the RKG metrics mentioning the number of individuals (i.e., 1809) in the KG, the number of axioms (i.e., 13578), and so forth.

| Metrics | |
|---|---|
| Axiom | 13578 |
| Logical axiom count | 10584 |
| Declaration axioms count | 2308 |
| Class count | 198 |
| Object property count | 220 |
| Data property count | 69 |
| Individual count | 1809 |
| Annotation Property count | 15 |

Fig. 13 - The RKG Metrics

Earlier (see Figure 8), we presented a visualization of the modelling aspect of the ontology through GraphDB[43]. The following Figure 14 shows Ranganathan at the central node linked to various other nodes (entities, such as *Sarada*, *Yogeshwar*, *Kaula*, *Palmer*, *Foskett*, *ISI*, *DRTC*, etc.) along with the associative relationships connected them (e.g., *inHonorOf*, *inspiredBy*). The partial RKG can be accessed at https://w3id.org/ontobio.

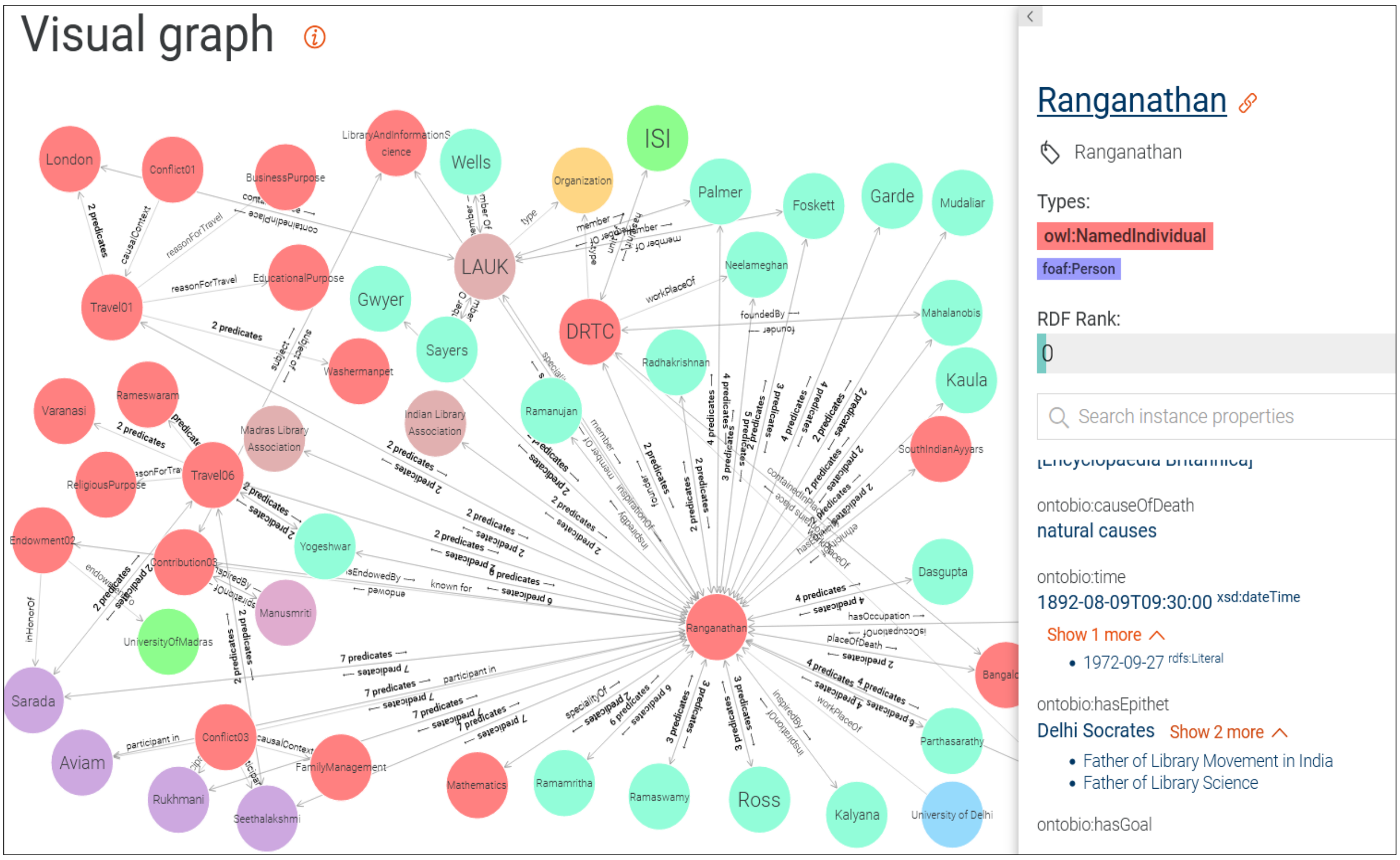

Fig. 14 - GraphDB Visualization of the Ranganathan KG

### *Evaluation of the Model*

The final phase focuses on evaluating the feasibility of the RKG—as to how well it aligns with its design objectives. The evaluation of RKG was conducted by running it through the Pellet reasoner and by executing a set of SPARQL queries. Reasoners are used for consistency checking, class membership, and classification of individuals. In particular, the Pellet reasoner[47] uses "a tableau-based decision procedure to provide many reasoning services (subsumption, satisfiability, classification, instance retrieval, conjunctive query answering) along with the capability to generate explanations for the inferences it computes". It did not return any critical or major issues with our model apart from some minor details.

The other way to evaluate the model was through SPARQL queries. As famously mentioned by Berners-Lee[48] "trying to use the Semantic Web without SPARQL is like trying to use a relational database without SQL". These queries help in matching graph patterns and retrieving those results that would answer our competency questions, although their application goes beyond their simple information retrieval purposes. Figures 15 & 16 present two examples of SPARQL queries using Apache Jena[49] based on the competency questions (which are discussed in the "Biographical Ontology Ontobio" section), along with the results.

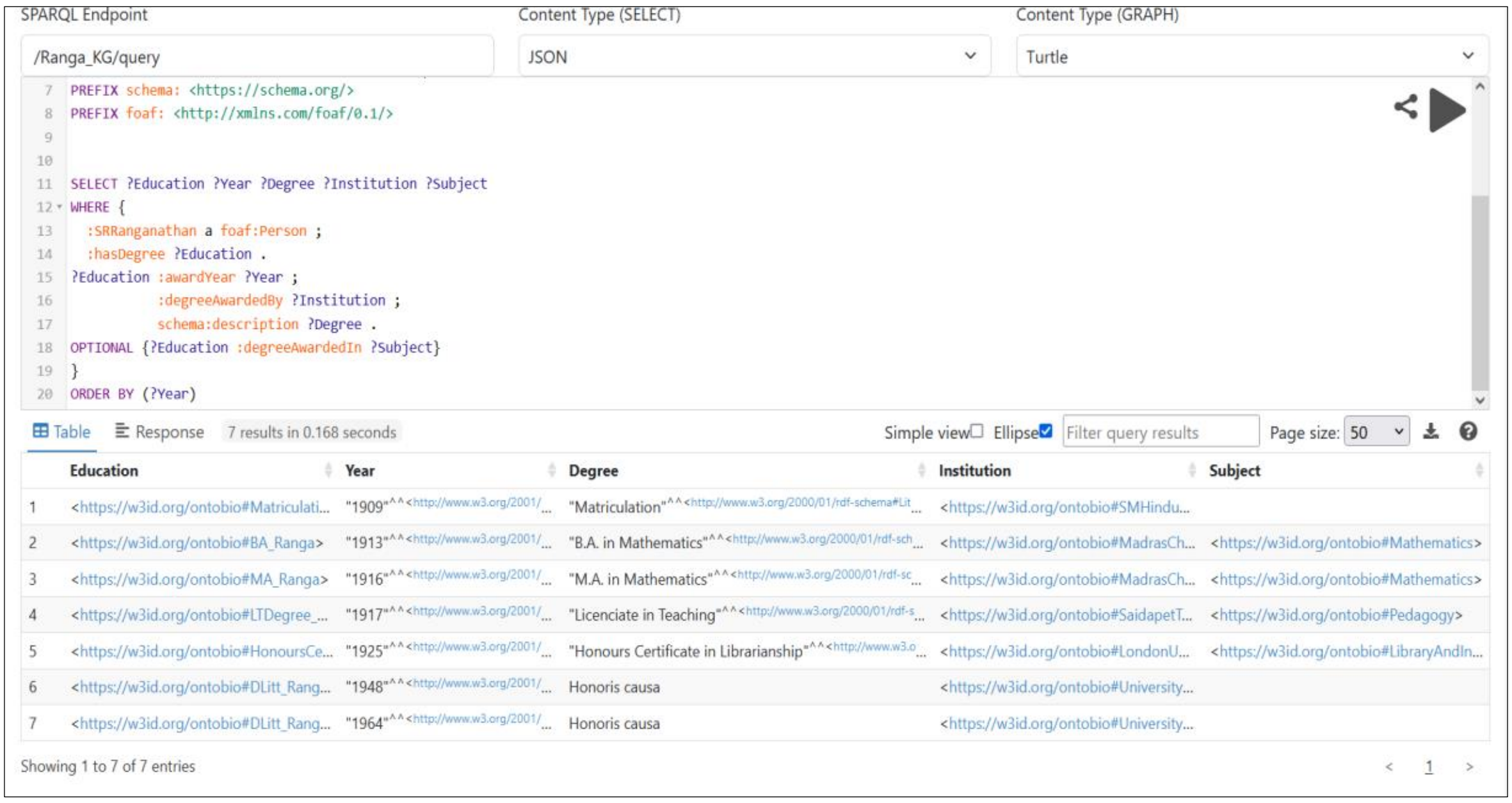

Fig. 15 - SPARQL Query to retrieve Ranganathan's educational details

The SPARQL query in Figure 15 is based on the third competency question (see "Biographical Ontology Ontobio" section) which retrieves Ranganathan's various educational details along with the degree name, awarding institution, and the date of award. The figure also depicts the query results. For example, he received his honorary doctorate degree (*honoris causa*) from the University of Pittsburgh[25] in 1964.

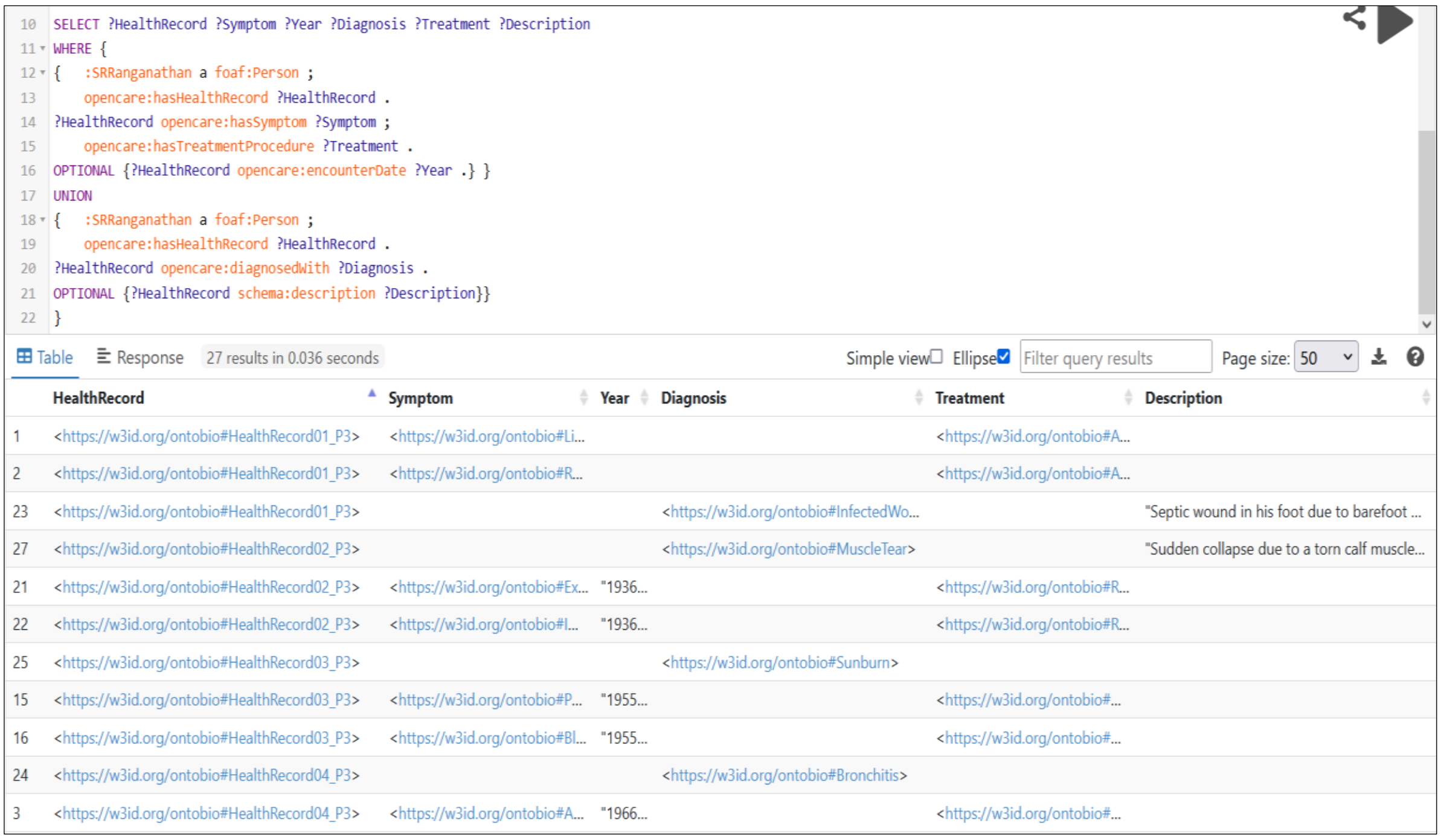

Fig. 16 - SPARQL Query for Ranganathan's health records using Apache Jena

The query shown in Figure 16 shows the major health issues that afflicted Ranganathan during his lifetime (query based on competency question no. 5). For example, while attending a wedding in 1936, Ranganathan suffered a muscle tear in his left leg resulting in extreme pain and immobility. He was ordered to rest and recover at home[26] for a while, medical advice that he eventually did not heed. We have five other health-related incidents in RKG which, due to space constraints, cannot be discussed here. Similarly, we ran many other queries successfully in our attempt to comprehensively cover the various aspects of Ranganathan's life and works.

## Conclusion and Discussion

Ranganathan has been hailed as "the greatest figure of librarianship in the 20th century[50]". His major ideas and theories continue to influence major technological advances in the twenty-first century[51]. As Weinberger notes: "It's as if Ranganathan designed his Colon Classification […] specifically for the kinds of instant reshuffling that were made feasible decades later in the Internet age […] Ranganathan was ahead of his time". It is fitting that we leverage a contemporary technological advance (in the form of KGs) to celebrate and enrich his life and works. We hope that our contribution bridges a noticeable gap in the Ranganathan legacy and serves as a collective call to action to further develop and strengthen his legacy[52]. We have already acknowledged major challenges in the form of out-of-print and hard-to-locate resources related to Ranganathan. It is something that can be achieved through a community effort and the dedication of academics, scholars, researchers, librarians, and students working on Ranganathan. We are aware of some of the limitations in the RKG with regards to the attributed quotes, anecdotes, dialogues, and multimedia assets that can only be incorporated into this KG with the assistance of other individuals. What we provide here is a substantial repository of information about Ranganathan, structured as graph data. As mentioned in the Introduction, KGs offer an intuitive understanding across various domains capturing the intricate relationships (through edges) among the entities (nodes) present in a given domain of interest; allowing for greater flexibility and the accommodation of heterogeneous data than it was possible through traditional methods[5]. Such a manifestation is well suited to collaborative work, the kind we envision necessary to achieve a fuller representation of Ranganathan's life and works.

We had indicated earlier the development of a live SPARQL endpoint (see the "Resources and Associated Challenges" section above) for anyone to access, query, and reuse the underlying data. Recognizing that most users are not proficient in writing SPARQL queries, we will provide a user-friendly demo and proper documentation of the methods online. It is worth noting that our current model does not encompass narrative elements such as anecdotes, dialogues, and stories related to Ranganathan. In a future effort, we aspire to include those aspects within a unified and integrative model. Lastly, as outlined in our motivations, we intend this dedicated KG to be the starting point for research related to Ranganathan and a tribute to his enduring contributions to the LIS profession.

## Acknowledgments

This work was carried out under the research project entitled "Facility for transforming library data to linked library data", funded by the Indian Council of Social Science Research (ICSSR), New Delhi.